\documentclass{article}
\begin{document}

\title{Interference, reduced action, and trajectories}

\author{Edward R.\ Floyd \\
10 Jamaica Village Road, Coronado, CA 92118-3208, USA \\
floyd@mailaps.org}

\date{30 July 2007}

\maketitle

\begin{abstract}
Instead of investigating the interference between two stationary, rectilinear wave
functions in a trajectory representation by examining the trajectories of the two
rectilinear wave functions individually, we examine a dichromatic wave function that
is synthesized from the two interfering wave functions. The physics of interference
is contained in the reduced action for the dichromatic wave function. As this
reduced action is a generator of the motion for the dichromatic wave function, it
determines the dichromatic wave function's trajectory.  The quantum effective mass
renders insight into the behavior of the trajectory.  The trajectory in turn renders
insight into quantum nonlocality.
\end{abstract}

\bigskip

\footnotesize

\noindent PACS Nos. 3.65Ta, 3.65Ca, 3.65Ud

\bigskip

\noindent Keywords: interference, trajectory representation, entanglement,
nonlocality, dwell time, determinism

\bigskip

\normalsize

\section{INTRODUCTION}

The trajectory representation of quantum mechanics is a nonlocal, phenomenological
theory that is deterministic. The quantum Hamilton-Jacobi equation underlies the
trajectory representation of quantum
mechanics.$^{(\ref{bib:prd34},\ref{bib:vigsym3})}$  The underlying Hamilton-Jacobi
formulation couches the trajectory representation of quantum mechanics in a
configuration space, time domain rather than a Hilbert space of wave mechanics.
Faraggi and Matone, using a quantum equivalence principle that connects all physical
systems by a coordinate transformation, have independently derived the quantum
stationary Hamilton-Jacobi equation (QSHJE) without using any axiomatic
interpretations of the the wave function, $\psi $.$^{(\ref{bib:fm},\ref{bib:fm2})}$
With Bertoldi, they have extended their work to higher dimensions and to
relativistic quantum mechanics.$^{(\ref{bib:bfm})}$ Without axiomatic
interpretations, the trajectory representation has been used to investigate the
foundations of quantum mechanics free of Copenhagen philosophy. The trajectory
representation has microstates that provides a counterexample showing that $\psi $
is not an exhaustive description of quantum
mechanics.$^{(\ref{bib:prd34},\ref{bib:fm2},\ref{bib:rc}-\ref{bib:fpl9})}$ The
trajectory interpretation has shown that the initial values of position and
momentum, of the Heisenberg uncertainty principle, form an insufficient subset of
initial values for determining the particular complete solution of the QSHJE:
acceleration and jerk, or their equivalents, must be included to form a necessary
and sufficient set of initial values to determine a unique
trajectory.$^{(\ref{bib:vigsym3},\ref{bib:fm2},\ref{bib:rc},\ref{bib:prd29})}$ The
trajectory representation also describes quantization
$^{(\ref{bib:prd34},\ref{bib:fm2},\ref{bib:prd25},\ref{bib:prd26})}$ and tunnelling
without probability.$^{(\ref{bib:fm2},\ref{bib:afb20})}$ Other aspects of the
trajectory representation have been studied.$^{(\ref{bib:misc},\ref{bib:bouda})}$

Interference is another fundamental attribute of quantum mechanics that
distinguishes it from classical mechanics.  Interference implies quantum
entanglement, which in turn implies nonlocality.  An initial assessment of
interference by the trajectory representation was noted as it incidently arose in
the study of transmission and reflection by a semi-infinite step
barrier.$^{(\ref{bib:pe7})}$ We examine interference in detail to gain new insight.
In the Copenhagen interpretation, the degree of interference between two plane wave
functions, $\psi_1(x)$ and $\psi_2(x)$ is manifested by the variation of Born's
probability density for the sum of entangled wave functions as a function of
position $x$, that is

\[
\bigl|[\psi_1(x)+\psi_2(x)]^{\dagger}[\psi_1(x)+\psi_2(x)]\bigr|.
\]

\noindent We examine herein the manifestation of interference in the deterministic
trajectory interpretation of quantum mechanics. The approach that we use considers a
solitary dichromatic particle whose spectral components interfere with each other
rather than studying the interference between one aggregate of particles and
another. In practice, we spectrally sum $\psi_1(x)$ and $\psi_2(x)$, each a
monochromatic wave function representing a spectral component, to form a dichromatic
wave function, $\psi_d(x)$, for a dichromatic particle.  From $\psi_d (x)$, we can
determine the dichromatic reduced action (Hamilton's characteristic function) and
trajectory. We find that trajectory representation describes destructive
interference and reinforcement as a dwell-time phenomenon of the trajectory weighted
by Faraggi and Matone's quantum equivalent mass,$^{(\ref{bib:fm2})}$ so invoking
probability and Born's probability density is unnecessary. We additionally find that
the trajectory representation describes phenomenon that has been attributed to
quantum-mechanical spreading of the wave function by the Copenhagen interpretation.
We also find that the consequent trajectories are nonlocal.

We investigate herein the prototype of interference problems, interference between
two stationary plane wave functions with the same wavelength.  This problem has been
examined by Holland using Bohmian mechanics.$^{(\ref{bib:holland})}$ The findings of
the trajectory representation differ with those of Bohmian mechanics as the two
representations have different equations of motion even though both representations
derive the same reduced action from the same
QSHJE.$^{(\ref{bib:vigsym3},\ref{bib:wyatt})}$

As part of this investigation, we show that while the trajectory representation has
analogies to group velocity of wave packets, the trajectory representation is more
general.

In a companion article, {\it welcher Weg} is examined for a simplified quantum
Young's diffraction experiment.$^{(\ref{bib:ww})}$

In Sect.\ 2, we investigate interference in one dimension in a trajectory
representation.  The interfering wave functions are used to synthesize a dichromatic
wave function. The dichromatic reduced action is developed as the generator of the
motion of interference patterns. Dwell times are developed for interference.
Nonlocal propagation is exhibited. In Sect.\ 3, we develop the contours of constant
reduced action in two dimensions. The evolution of the reduced action as it goes
from representing a running wave function through a dichromatic wave function to a
standing wave function is developed. Interference is shown to cause the trajectory
not to be orthogonal to the contours of reduced action in general. In Sect.\ 4, the
conclusions are summarized.

\section{EQUATION OF MOTION}

Let us first examine the simple case of interference between two rectilinear,
monochromatic wave functions of equal wavelength in one dimension.  One wave
function, $\psi_+(x) = A \exp(ikx)$, propagates in the positive $x$-direction; the
other,$\psi_-(x) = B \exp(-ikx -i\beta)$, in the negative $x$ direction.  The
amplitudes $A$ and $B$ are real, non-negative constants, $+k$ and $-k$ are
respectively the wave numbers for the interfering wave functions $\psi_+$ and
$\psi_-$, and $\beta$ is a constant phase shift in $\psi_-$. Hence we only need one
dimension to investigate this interference.  We arbitrarily assume that $A>0$ and $A
\ge B\ge 0$. Alternatively, this interference may be considered to be the
interference between a running wave function, $(A-B) \exp(ikx)$, and a standing wave
function, $2B \exp(-i\beta/2) \cos(kr+\beta/2)$.

We may spectrally synthesize a dichromatic wave function, $\psi_d(x)$, with only two
spectral components, $+k$ and $-k$. The subscript ``$d$" in any function, for
example $f_d(x)$, will designate that this function describes some property of the
dichromatic wave function.  The dichromatic wave function is a time-independent
solution of the Schr\"odinger equation by the superpositional principle of linear
homogeneous differential equations.  As such, $\psi_d (x)$ is the wave function for
the dichromatic particle.  Historically in quantum mechanics, the dichromatic wave
function was first developed by a trajectory representation in Ref. \ref{bib:pe5}
(where it was identified as a compoundly modulated wave) rather then by a
Schr\"{o}dinger representation. The complete solution for the reduced action of the
QSHJE rendered compoundly modulated (dichromatic) wave
functions.$^{(\ref{bib:pe5})}$ The dichromatic wave function is described
by$^{(\ref{bib:holland},\ref{bib:pe5})}$

\begin{eqnarray}
\psi_d(x)& = & \psi_+(x) + \psi_-(x) \nonumber \\
         & = & [A^2+B^2+2AB\cos(2kx+\beta)]^{1/2} \exp \left[ i \arctan \left(
           \frac{A \sin(kx) - B \sin(kx+\beta)}{A \cos(kx) + B \cos(kx+\beta)}\right) \right].
\label{eq:psi}
\end{eqnarray}

\noindent  The two running wave functions, $\psi_+$ and $\psi_-$, are now spectral
components, $+k$ and $-k$, of $\psi_d$.  The dichromatic wave function, $\psi_d$,
innately incorporates the interference between its two spectral components. We
consider that our particle of interest is a dichromatic particle whose wave function
is $\psi_d(x)$.  In the limit $A \to B$, then the last line of Eq.\ (\ref{eq:psi})
still renders $\lim_{A \to B} \psi_d(x) = 2B \cos(kx), i2B \sin(kx)$ for
$\beta=0,\pi$ respectively. Destructive and constructive interference is manifested
by the cosine term in the amplitude of the dichromatic wave function $\psi_d $ in
Eq.\ (\ref{eq:psi}). The Copenhagen interpretation of quantum mechanics attributes
Born's probability density to the absolute value of $\psi_d^{\dagger}(x) \psi_d(x)$.
Note that $\psi_+,\ \psi_-$ and $\psi_d$ are all solutions of the time-independent
Schr\"{o}dinger equation for the free particle,

\[
-\frac{\hbar^2}{2m}\frac{\partial^2 \psi }{\partial x^2} = \frac{\hbar^2k^2}{2m}\psi
\]

\noindent where $\hbar$ is Planck's constant and $m$ is particle mass.  The energy,
$E$, for the free dichromatic particle is specified by $E = \hbar^2k^2/(2m)$.

We recognize that there is a degree of arbitrariness in specifying that the
dichromatic wave function, $\psi_d$ is synthesized by the interference of two
running wave functions, $\psi_+$ and $\psi_-$. As another dichromatic wave function,

\[
\psi_{dd} = [A^2+B^2-2AB\cos(2kx+\beta)]^{1/2} \exp \left[ i \arctan \left(
           \frac{A \sin(kx) + B \sin(kx+\beta)}{A \cos(kx) - B \cos(kx+\beta)}\right) \right],
\]

\noindent is also a solution of the Schr\"{o}dinger equation and is, for finite
$A/B$, independent of $\psi_d$, then $\psi_+ = \psi_d + \psi_{dd}$. By the
superpositional principle of linear homogeneous differential equations, one can
always specify that the running wave function $\psi_+$ is formed by the interference
between two dichromatic wave functions, $\psi_d$ and $\psi_{dd}$.  All sets of
independent solutions of the time-independent Schr\"{o}dinger equation are equally
valid.  For this investigation, we chose one solution to be $\psi_d(x)$ for
heuristic purposes.

The reduced action $W_d$ for the dichromatic wave function is given by

\begin{equation}
W_d = \hbar \arctan \left( \frac{A \sin(kx) - B \sin(kx+\beta)}{A \cos(kx) +
B\cos(kx+\beta)} \right). \label{eq:w}
\end{equation}

\noindent  Likewise, had we been interested in free particles with one spectral
component, then the reduced actions for $\psi_+$ would be given by $W_+ = +\hbar
kx$; for $\psi_-$, by $W_- = -\hbar kx$. Note that $W_d,\ W_+$ and $W_-$ are all
solutions of the QSHJE for the free particle

\begin{equation}
\frac{(\partial W/\partial x)^2}{2m} - \frac{\hbar^2k^2}{2m} +
\underbrace{\frac{\hbar^2}{4m} \left[ \frac{\partial ^3 W/\partial x^3}{\partial
W/\partial x} - \frac{3}{2} \left( \frac{\partial ^2 W/\partial x^2}{\partial
W/\partial x} \right)^2 \right]}_{{\mbox{\footnotesize Bohm's quantum potential,\ }}
Q} = 0. \label{eq:QSHJE}
\end{equation}

\noindent We shall demonstrate the trajectory representation of interference by
investigating the reduced action, $W_d$, for the dichromatic wave function as
exhibited by Eq.\ (\ref{eq:w}).  Interference between the two spectral components of
$\psi_d$ is innate in $W_d$. The reduced action, $W_d$, is the generator of the
motion for manifesting interference and the dynamics of the dichromatic wave
function.

The conjugate momentum, $p_d$, for the dichromatic wave function is defined as the
gradient of the reduced action. For the dichromatic wave function in one dimension,
the conjugate momentum is

\begin{equation}
p_d \equiv \frac{\partial W_d}{\partial x} =\frac{\hbar k (A^2-B^2)}{A^2+B^2+2AB
\cos(2kx+\beta)}. \label{eq:cm}
\end{equation}

\noindent  The interference is manifested in the conjugate momentum by the cosine
term in the denominator of Eq.\ (\ref{eq:cm}).  The conjugate momentum of the
dichromatic wave function is by Eq.\ (\ref{eq:psi}) inversely proportional to the
inverse square of amplitude of the dichromatic wave function.  Hence, Born's
probability density, $\rho_d(x)$  is related to the conjugate momentum by Eqs.\
(\ref{eq:psi}) and (\ref{eq:cm}) as

\begin{equation}
\rho_d(x) = \psi_d ^{\dagger}(x) \psi_d (x)  \prec [(\partial W_d/\partial
x^{\prime})^{-1}]|_{x^{\prime} = x}, \label{eq:rho}
\end{equation}

\noindent  for the dichromatic wave function. Thus, the conjugate momentum for the
dichromatic wave function provides an alternative explanation of interfering wave
functions without any need to invoke Born's probability density. For completeness,
we note that for bound states where $\psi$ is real, then Eq.\ (\ref{eq:rho}) does
not hold.$^{(\ref{bib:fpl9})}$ This will be discussed further later.

We use Jacobi's theorem to determine the quantum equation of
motion.$^{(\ref{bib:prd34},\ref{bib:vigsym3},\ref{bib:fm2},\ref{bib:rc})}$ As
Jacobi's theorem also determines the equation of motion in classical mechanics,
Jacobi's theorem is universal transcending across the division between classical and
quantum mechanics. In the classical limit, trajectories are consistently determined.
Note that this is a departure from Bohmian mechanics, which would have assumed that
$p_d$ be the mechanical momentum $m \dot{x}_d$, from which the Bohmian quantum
equation of motion be determined by integrating
$p_d$.$^{(\ref{bib:fm2},\ref{bib:prd26})}$ For comparison, the Bohmian trajectory
for interference between plane wave functions has been given by
Holland.$^{(\ref{bib:holland})}$ Jacobi's theorem renders

\begin{equation}
t_d-\tau = \partial W_d/\partial E =\frac{mx(A^2-B^2)}{\hbar k[A^2+B^2 +2AB
\cos(2kx+\beta)]} \label{eq:eom}
\end{equation}

\noindent where $\tau $ is the Hamilton-Jacobi constant coordinate (a nontrivial
constant of integration) in units of time necessary for determining the motion of
the dichromatic wave function.

Faraggi and Matone have shown that $\dot{x}_d$ may be expressed from Eq.\
(\ref{eq:eom}) by$^{(\ref{bib:fm2})}$

\begin{equation}
\dot{x}_d = \frac{1}{\partial t_d/\partial x} = \frac{1}{\partial^2 W_d/\partial x
\partial E} = \frac{1}{\partial^2 W_d/\partial E \partial x}=\frac{1}{\partial
p_d/\partial E}. \label{eq:vel}
\end{equation}

\noindent We designate $\partial t_d/\partial x$ to be dwell density where
$\int_{x_1}^{x_2} \partial t_d/\partial x \, dx = t_d(x_2) - t_d(x_1)$ renders the
dwell time that the dichromatic particle spends in the domain between $x_1$ and
$x_2$. Before we evaluate the velocity, we present a brief digression. Let us now
associate the constant $E$ with some time-independent quantum Hamiltonian ${\cal
H}(x,p)$ such that ${\cal H}=E$ while ${\cal H}$ still obeying the Faraggi-Matone
quantum equivalence principle (the subscript $d$ is dropped in this digression). A
velocity, analogous to Eq.\ (\ref{eq:vel}), may then be expressed as

\[
\dot{x}= \frac{\partial {\cal H}}{\partial p}
\]

\noindent which is half of the canonical equations of Hamilton for time independence
where it is noted that $p$ is the conjugate momentum and not the mechanical
momentum. The other half follows from ${\cal H}(x,p)$ being a constant and is given
by

\[
\dot{p}= -\frac{\partial {\cal H}}{\partial x}.
\]

\noindent This digression has been presented for heuristic purposes only.  It shows
that quantum Hamilton-Jacobi representation would be consistent with the the
canonical equations for a quantum ${\cal H}(x,p)$ analogous to classical mechanics.
Regrettably, we presently do not know the algorithm to develop ${\cal H}(x,p)$. The
problem is that a reversible canonical transformation is needed to generate the
classical Hamilton's principal function from the classical Hamiltonian,
$H(x,p_{\mbox{\scriptsize mech}})$ where $p_{\mbox{\scriptsize mech}}$ is the
mechanical momentum, $p_{\mbox{\scriptsize mech}}=m \dot{x}$. On the other hand, the
Faraggi-Matone quantum equivalence principle is restricted to just coordinate
transformations.$^{(\ref{bib:fm2})}$  The interested reader who desires to know more
on applying trajectory representation of quantum mechanics in a non-Hamilton-Jacobi
formulation is referred to Bouda et al$^{(\ref{bib:bouda})}$ for progress in
``quantum analytical mechanics", to Wyatt$^{(\ref{bib:wyatt})}$ for progress in
``quantum trajectory methods", and to Poirier$^{(\ref{bib:poirier})}$ for progress
in ``counterpropagating wave method".

We now return to Eq.\ (\ref{eq:vel})and evaluate the velocity, $\dot{x}_d$, for the
dichromatic wave function as

\begin{equation}
\dot{x}_d=(\partial t/\partial x_d)^{-1}=\frac{\hbar k}{m(A^2-B^2)} \frac{[A^2+B^2 +
2AB \cos(2kx+\beta)]^2}{A^2+B^2 + 2AB \cos(2kx+\beta) + 4 AB kx \sin(2kx+\beta)}.
\label{eq:velocity}
\end{equation}

\noindent From Eqs.\ (\ref{eq:cm}) and (\ref{eq:velocity}), the conjugate momentum
for the dichromatic wave function is not the mechanical momentum, i.e., $p_d \ne m
\dot{x}_d$.

Faraggi and Matone introduced an ``effective quantum mass", $m_Q$, such that $m_Q
\dot{x}=\partial W/\partial x$ or $m_Q=m(1-\partial Q/\partial E)$ where $Q$ is
Bohm's quantum potential in Eq.\ ({\ref{eq:QSHJE}).$^{(\ref{bib:fm2})}$  Faraggi and
Matone were inspired by special relativity to introduce $m_Q$.  Solid state physics
provides a precedence where the effective mass, $m^{\ast}$, for electrons and holes
accounts for the interaction between an electron or hole and a crystal as the
electron or hole responds to an external force. In the band theory of solids,
$m^{\ast}$ may be positive, negative, zero or infinite.

For the dichromatic wave function, the effective quantum mass, $m_{Q_d}$ is given by

\begin{equation}
m_{Q_d} = m \frac{(A^2-B^2)^2 [A^2+B^2+2AB \cos(2kx+\beta)+4ABkx
\sin(2kx+\beta)]}{[A^2+B^2+2AB \cos(2kx+\beta)]^3}. \label{eq:eqm}
\end{equation}

For the dichromatic wave function by Eq.\ (\ref{eq:rho}), the Born's probability
density, $\rho_d$ is proportional to dwell density $\partial t_d/\partial x$ divided
by $m_{Q_d}$ as

\begin{equation}
\rho_d = \psi_d ^{\dagger}(x) \psi_d (x) \prec \frac{1}{\partial W_d/\partial x} =
 \frac{1}{m_{Q_d} \dot{x}} \prec {\partial t/\partial x}.
\end{equation}

\noindent Thus, the trajectory representation of interference need not attribute a
probability amplitude to the dichromatic wave function $\psi_d(x)$.

For completeness, we note that if the wave function is real, then $\psi
^{\dagger}(x) \psi (x) \ne (\partial W/\partial x)^{-1}$.  Elsewhere, the trajectory
interpretation has used the concept of dwell time to describe phenomena for real
$\psi (x)$.  The trajectory representation predicts the dwell times for
tunnelling$^{(\ref{bib:afb20})}$ and reflection$^{(\ref{bib:fpl13})}$ that are
consistent with those of Barton$^{(\ref{bib:barton})}$,
Hartman$^{(\ref{bib:hartman})}$, and Fletcher.$^{(\ref{bib:fletcher})}$ Also, the
trajectories of bound state square wells at their turning points at $x=\pm \infty$
manifest that $\dot{x} \to \pm \infty$ while $\partial W/\partial x \to 0$ implying
that the trajectory deep in the forbidden region transits an infinite distance in a
finite time.$^{(\ref{bib:fpl13})}$ A suggested generalization for the effective
quantum mass is

\[
m_Q = \frac{\partial t/\partial x}{\psi ^{\dagger}(x) \psi (x)}
\]

\noindent for real or complex $\psi (x)$.

Let us now investigate a particular case.  We consider the case that $\hbar=1,\
m=1,\ k=\pi/2, A=1,\ B= 0.5,\ \beta=0,\pi$ and $\tau=0$. The trajectories for
$\beta=0,\pi$ are exhibited on Fig.\ 1 where the trajectories for the dichromatic
wave function exhibit quasi-periodic reversals in time where $\partial t_d/\partial
x = 0$. These time reversals occur in the vicinity of $2kx+\beta = n \pi,\ n=1,2,
\cdots$.  The time reversals induce nonlocality on two counts. First, the time
reversals as extremum in time induce an instantaneous infinite velocities in
accordance with Eq.\ (\ref{eq:vel}). Second, the time reversals imply pair creations
and annihilations of interference patterns. For $B/A \ll 1$, latent early time
reversals may be suppressed. The upper time reversals manifest destructive
interference where the absolute value $\psi_d$ form a local minima. Likewise, the
lower time reversal manifest constructive interference. By Eq.\ (\ref{eq:eom}) for
$A=1$ and $B=0.5$, the trajectories in Fig.\ 1 have the locus of their upper time
reversals, $t_u$ on a straight line in the $t,x$-plane given by

\begin{equation}
t_u = \frac{A+B}{A-B} \frac{mx}{\hbar k} \label{eq:upper}
\end{equation}

\noindent  for any phase shift $\beta$.   Likewise, the locus of lower time
reversals, $t_{\ell}$, that manifest constructive interference (reinforcement) are
located along the straight line in the $t,x$-plane given by

\begin{equation}
t_{\ell} = \frac{A-B}{A+B} \frac{mx}{\hbar k}  \label{eq:lower}
\end{equation}

\noindent  for any phase shift $\beta$.  The lines $t_u(x)$ and $t_{\ell}(x)$ form
wedge that is densely covered by trajectories launched from origin of Fig.\ 1 by
adjusting the phase factor $\beta$. As the trajectories are horizontal (parallel to
the $t$-axis) at these reversal points, the lines $t_u$ and $t_{\ell}$ are not
caustics of the trajectories for finite $A$ and $B$ and $A \ne B$.  As illustrated
by Fig.\ 2, additional trajectories for $\beta = \pi/4, \pi/2, \cdots, 7\pi/4$ show
the forming of caustics near the lines of time reversal but outside the wedge formed
by $t_u$ and $t_{\ell}$. The area in the $t,x$-plane between the caustics is
designated the extended wedge. The time reversals of the trajectories ensure that a
solitary trajectory continues to span the extended wedge quasi-periodically.

While the trajectories for the dichromatic wave function have time reversals, this
does not imply that the dichromatic wave function itself reverses direction. Readers
interested in the phenomenon of direction reversing travelling (running) waves are
referred to Knobloch et al.$^{(\ref{bib:knobloch})}$

In order to gain insight, we now suspend our investigation of the time reversal
phenomenon exhibited by Fig.\ 1 until after we consider two limiting cases: first,
the running wave function only case (i.e., $B=0$); and, second, the standing wave
function only case (i.e., A=B=1 for a cosine wave function). For the first limiting
case, if we let $B \to 0$ for our particular case, then $\psi_d = A \exp(ikx)$ as
the dichromatic wave function becomes a pure running wave function only. As $B \to
0$, then $t_u \to t_{\ell}$ by Eqs.\ (\ref{eq:upper}) and (\ref{eq:lower}).  The
equation of motion, Eq.\ (\ref{eq:eom}) becomes $t_d=mx/(\hbar k)$ eliminating all
time reversals. The velocity becomes $\dot{x}_d=\hbar x/m$, and the effective
quantum mass becomes $m_{Q_d}=m$ as expected.

For the second limiting case, we let $A=1=B+\epsilon$ where $\epsilon$ is real and
nonnegative so that as $\epsilon \to 0$, then $B$ approaches $A$ from below.  For
$\lim _{\epsilon \to 0} \psi _d =2A \cos(kx)$ and the two edges of the wedge given
by Eqs.\ (\ref{eq:upper}) and ({\ref{eq:lower}) become orthogonal to each other so
that the wedge spans the entire quadrant $t,x \ge 0$ of $t,x$-plane in Fig.\ 1 as
the dichromatic wave function becomes entirely a standing wave function. The
equation of motion, Eq.\ (\ref{eq:eom}), for this standing wave function would then
become

\begin{equation}
\lim _{\epsilon \to 0} t_d  =  \sum _{n=1}^{\infty}\{ \delta [x-(2n-1)\pi /(2k)] +
\delta[x+(2n-1)\pi /2k] \} \label{eq:lpneg}
\end{equation}

\noindent where $\delta$ is the Dirac delta function.  Equation (\ref{eq:lpneg})
presents the motion for a standing wave function whose launch point is at
$x=-\infty$ with the constant coordinate $\tau=0$. But the standing wave function
whose launch point is at $x=0$ is the more interesting case to examine. Thus, we
shall restrict Eq.\ (\ref{eq:lpneg}) to the domain $x>0$ where Eq.\ (\ref{eq:lpneg})
becomes

\begin{equation}
\lim _{\epsilon \to 0} t_d = \sum _{n=1}^{\infty}\delta [x-(2n-1)\pi /(2k)] = \sum
_{n=1}^{\infty}\{ \delta [x-(2n-1)], \ \ x>0. \label{eom+}
\end{equation}

\noindent In the domain $x<0$, we now relax the requirement that $B<A$. We use the
subscript $-d$ to denote a property of the dichromatic wave function for with $B>A$.
In the alternative representation, $\psi_{-d}$ represents the interference between a
running wave function in the negative $x$-direction, $(B-A)\exp(-ikx)$ and a
standing wave function $2A \cos(-kx)$.  The corresponding reduced action is still
$W_{-d} = \hbar \arctan \left( \frac{A-B}{A+B} \tan(kx)\right)$ consistent with Eq.\
(\ref{eq:w}). The equation of motion is still rendered by Eq.\ (\ref{eq:eom}), but
since $B>A$, motion is in the negative $x$-direction.   We let $A=B-\epsilon$ where
$\epsilon$ is real and nonnegative so that as $\epsilon \to 0$, then $B$ approaches
$A$ from above.  The equation of motion, Eq.\ (\ref{eq:eom}), for this standing
motion in the domain $x<0$ is

\begin{equation}
\lim _{\epsilon \to 0} t_{-d} = -\sum _{n=1}^{\infty} \delta [x-(2n-1)\pi /(2k)] =
-\sum _{n=1}^{\infty}\{ \delta [x-(2n-1)], \ \ x<0. \label{eom-}
\end{equation}

\noindent Thus, the standing wave function with launch point at $x=0$ has positive
infinite velocity for $x>0$ and negative infinite velocity of propagation for $x<0$
except at the nulls, $x=\pm 1,\pm 2,\pm 3,\cdots$, of the the standing wave
function, $2 \cos(kx)$, where the null innately has nil velocity of propagation.
This is consistent with the concept of a standing wave function. The effective
quantum mass would become

\begin{eqnarray*}
\lim_{\epsilon \to 0} m_{Q_{\pm d}} & = & 0, \ \ x \ne \pm 1,\pm 3,\pm 5, \cdots \\
                       & = & \infty, \ \ x = \pm 1,\pm 3,\pm 5, \cdots.
\end{eqnarray*}

\noindent  We note that the effective quantum mass becomes infinite where the
velocity of the standing wave function is nil and becomes nil where the velocity is
infinite.

We now return to investigating the particular case exhibited in Fig.\ 1 where $A=1$,
$B=0.5$, $k=\pi/2$ and $\beta = 0$. The motion of propagation of the dichromatic
wave function, $\psi_d (x)$ is a combination of the running wave function
$(A-B)\exp(ikx)=2^{-1}\exp(i \pi x/2)$ and the standing wave function $2B \cos(kx) =
\cos(\pi x/2)$. As $B$ becomes smaller for $B/A \le 1$, the wedge between the upper
and lower turning points, as given by Eqs.\ (\ref{eq:upper}) and (\ref{eq:lower}),
narrows as the motion becomes more like a running wave function. Conversely, as $B$
increases for $B/A \le 1$, the allowed wedge of propagation widens manifesting the
increasing influence of the standing wave function.

Another way of examining the propagation implied by Fig.\ 1 uses the effective
quantum mass, $m_Q$, of Faraggi and Matone, Eq.\ (\ref{eq:eqm}) and the velocity,
$\dot{x}_d$ of Eq.\ (\ref{eq:velocity}). Both $m_{Q_d}$ and $\dot{x}_d$ are poorly
behaved for our particular case, so we tame their behavior with the Euler-inspired
transforms

\[
{\cal M}_{Q_d} = 0.8\ \mbox{sgn}(m_{Q_d}) \frac{m_{Q_d}^2}{1+m_{Q_d}^2} \ \
\mbox{and}  \ \ \stackrel{\bullet}{{\cal X}}_d = \mbox{sgn}(\dot{x}_d)
\frac{\dot{x}_d^2}{1+\dot{x}_d^2}.
\]

\noindent  We plot the transformed effective quantum mass, ${\cal M}_{Q_d}$, and the
transformed velocity, $\stackrel{\bullet}{{\cal X}}_d$, on Fig.\ 3 as function of
$x$. For clarity of exposition, ${\cal M}_{Q_d}$ is normalized differently than
$\stackrel{\bullet}{{\cal X}}_d$.   As exhibited by Fig.\ 3, the effective quantum
mass, $m_{Q_d}$ for the dichromatic wave function offers insight into the
nonlocality at the time reversal points. The effective mass, $m_{Q_d}$ is negative
wherever the trajectory exhibits retrograde motion with respect to time. The
effective quantum mass has the property that $m_{Q_d} = 0$ for the dichromatic wave
function wherever $\dot{x}_d \to \pm \infty$ at the extrema in time denoting the
time reversal points where trajectory segments with $m_{Q_d}$ of opposite sign are
either created or annihilated.  The local extrema in time at the turning points
induce an instantaneous infinite velocity. Still, the nil quantum effective mass for
the dichromatic wave function at the turning points implies that pair creation would
not be an endoergic process and that pair annihilation likewise would not be an
exoergic process. Hence, neither creation nor annihilation of pairs of interference
patterns implies a high energy process, which is consistent with $\psi_d$ and $p_d$
being solutions for energy $E=\hbar^2k^2/(2m)$ to the Schr\"{o}dinger equation and
QSHJE respectively. For completeness, Fig.\ 3 is only quasi-periodic.

As exhibited by Fig.\ 1, the trajectory segments are annihilated near
$x=1,3,5,\cdots$ where by Eq.\ (\ref{eq:w}), the reduced action for the particular
case becomes $ W_d = \arctan [3^{-1} \tan(x \pi/2)] $ inducing jumps to new Riemann
sheets at $x=1,3,5,\cdots$. The annihilations just manifest the confluence of two
interference patterns that each have reached their point of maximum destructive
interference. Likewise, the creation of two trajectory segments near
$x=2,4,6,\cdots$ manifests the launching of two interference patterns travelling in
opposite directions --- one pattern in the advancing direction while the other in
retrograde with respect to time. As the interference patterns travel from their
lower turning points to the upper turning points, the interference goes
monotonically from constructive to destructive.

The study of the group velocity$^{(\ref{bib:bw})}$ of wave packets in a general wave
propagation renders a precedent. The group velocity is given by the partial
derivative of frequency with respect to wavelength is reminiscent of Jacobi's
theorem.  Group velocity is only a valid concept as long as the wave packet
maintains its integrity. Any creation or annihilation of a pair of interference
patterns are sufficient to disrupt the integrity of the wave packet. In the general
theory of wave packet propagation, integrity is maintained for a finite duration by
having a carrier frequency that is very large in comparison to the frequency
modulation.  The trajectory representation renders a generalization of the group
velocity on two counts.  First, as the dichromatic wave function, $\psi_d$, given by
Eq.\ (\ref{eq:psi}) has no carrier wave, the trajectory representation for the
dichromatic wave function has generalized the group velocity concept to cover
situations without a carrier wave. Second, the trajectory representation relaxes the
requirement that the dichromatic wave function have spectral components with
approximately equal strength (this relaxation can be extended to polychromatic wave
functions). As the integrity of interference patterns for the individual trajectory
segments for dichromatic wave function is comparatively short in the trajectory
representation, the transition to other trajectory segments, whose motion is in the
opposite direction, is accomplished by pair creations and annihilations of
interference patterns at the turning points where the quantum effective mass goes to
zero. The trajectory representation thus extends trajectories beyond the point where
the wave packet loses its integrity.

There is an alternative explanation to the loss of integrity of the wave packet and
the generation of pair creations and annihilations of interference patterns.  Any
finite self-entanglement innately prescribes that as the dichromatic wave function
propagates away from it launch point at $x_0$ along its trajectory, it eventually
reaches a point where the interfering components, $\psi_1=A\exp[ik(x-x_0)]$ and
$\psi_2=B\exp[-ik(x+x_0)]$, have separated sufficiently far, $2x$, to induce the
loss of wave packet integrity. There, self-entanglement compels the nonlocal
generation of pair creation and annihilations of interference patterns.  In this
manner, self-entanglement, for which nonlocality is inherent as $\psi_d \ne K \psi_1
\psi_2$ where $K$ is some constant, manifests strong nonlocality for the dichromatic
particle can be at several separated locations simultaneously.  The underlying
physics for this strong nonlocality is that the dichromatic particle travels between
particular points $x_1$ and $x_2$ in one-dimensional space along a series of
trajectory segments that alternate between forward and retrograde motion with
respect to time such that the transit time between $x_1$ and $x_2$ is nil. As the
dichromatic particle may be simultaneously in multiple locations, the nonlocal
dichromatic particle is a distributed particle, and its trajectory manifests motion
for a distributed particle.

For completeness, we present a Copenhagen interpretation of quantum mechanics of
this Section.  Copenhagen rejects determinism and trajectories out of hand but does
accept the concept of wave packets.  By Copenhagen, the wedge of allowed
trajectories in Fig.\ 1 would manifest that, first, the wave packet of the
dichromatic wave function was launched at the apex of the extended wedge at $x=0$
and $t=0$.  Second, by Copenhagen, as this wave packet subsequently propagates away
from the apex of the extended wedge, the wave packet would spread with time as a
function of the broadness of the wedge.  Third, this wave packet by Copenhagen would
eventually spread so far that it would loose its integrity where the trajectory
representation manifests a time reversal (retrograde motion). However, by
Copenhagen, loss of wave packet integrity would not preclude further spreading.
Copenhagen would also attribute the measurement of $x$ at time $t$ to a collapse of
the dichromatic wave function to the position $x$ whose distribution of measurement
positions would manifest only a Born probability density,
$\psi_d^{\dagger}(x)\psi_d(x)$, and not some underlying trajectory of the
dichromatic wave function.  Finally, Copenhagen would agree that the dichromatic
wave function, $\psi_d(x)$, would be an entangled combination of $\psi_1(x)$ and
$\psi_2(x)$ as $\psi_d(x)$ has already been shown to be not factorable into
$\psi_1(x)$ and $\psi_2(x)$.

\section{REDUCED ACTION AND TRAJECTORY EQUATION}

In this Section, we exhibit in two dimensions the contours of reduced action that
represent interference and develop the trajectory equation. Let us consider the
reduced action for a wave function given by

\begin{eqnarray}
\psi(x,y) & = & [A\exp(ik_xx)+B\exp(-ik_xx)]\exp{(ik_yy)} \nonumber \\
         & = & [A^2+B^2+2AB\cos(2k_xx)]^{1/2}
        \exp\left[i\arctan\left(\frac{A-B}{A+B}\tan(k_xx)\right) +ik_yy\right], \label{eq:psi2}
\end{eqnarray}

\noindent which is separable.  The wave function< $\psi(x,y)$, is a solution of the
Schr\"{o}dinger equation for the free particle of energy $E = \hbar^2(k_x^2 +
k_y^2)/(2m)$. The right side of the first line of Eq.\ (\ref{eq:psi2}) represents
two wave functions, each also a solution to the same Shr\"{o}dinger equation, that
interfere with each other. The right side of the second line of Eq.\ (\ref{eq:psi2})
is a wave function and with innate internal interference and has been synthesized
from the two wave functions of the right side of the first line of Eq.\
(\ref{eq:psi2}). This synthesized wave function is denoted herein as $\psi_i$
(dichromatic wave is no longer proper terminology for the spectral component $k_y$
adds a third spectral component). The reduced action, $W_i$, for $\psi_i$ is given
by

\begin{equation}
W_i = \hbar \left[ \arctan\left(\frac{A-B}{A+B}\tan(k_xx)\right)  + k_yy\right]
\label{eq:w2}
\end{equation}

\noindent where $y$ is a cyclic coordinate.  The trajectory equation which renders
the constant coordinate $y_0$ in accordance with Jacobi's theorem is given by

\begin{equation}
y_0 = \frac{\partial W}{\partial k_y} = y - \frac{(A^2-B^2)(k_y x)/k_x}{A^2 + B^2 +
2AB \cos(2k_x x)}. \label{eq:te}
\end{equation}

Let us consider a particular case.  We let $A=1$, $B=0.5$, $\hbar = 1$, $m = 1$ and
$k_x = k_y = \pi/2$. We present selected contours of constant reduced action, $W$ of
Eq.\ (\ref{eq:w2}), as solid lines in Fig.\ 4. These selected contours are separated
by $h/4$ units of action (we have set $\hbar=1)$.  Due to the periodic
$x$-dependence on the right side of Eq.\ (\ref{eq:w2}), the contours of constant
reduced action diagonally serpentine across the infinite $x,y$-plane. We also
present on Fig.\ 4 the trajectory whose launch point is the origin ($y_0=0$) as a
dashed line by applying Eq.\ (\ref{eq:te}) to this particular case.

We now consider the more general case where the coefficient $B$ is replaced by
$B\exp(i\beta)$, which introduces a phase shift $\beta$.  Analogous to Eqs.\
(\ref{eq:upper}) and (\ref{eq:lower}), the trajectory has upper and lower turning
points in the $y$-direction, $y_u$ and $y_{\ell}$ respectively, whose loci for $0
\le \beta \le 2\pi$ are given by

\[
y_u = \frac{A+B}{A-B} k_y x/k_x \ \ \ \ \mbox{and} \ \ \ \ y_{\ell} =
\frac{A-B}{A+B} k_y x/k_x.
\]

\noindent While not shown, the set of trajectories over range of phase shifts $0 \le
\beta \le 2\pi$ would form caustics near the loci of $y_u$ and $y_{\ell}$ analogous
to the caustics of Fig.\ 2. Note that in general the trajectory is not orthogonal to
the contours of constant reduced action in Fig.\ 4 in contrast to Bohmian mechanics.
In particular, the trajectory becomes tangent to a surface of constant reduced
action at some point near the trajectory's turning point. There is a precedent for
this as the trajectories for oblique reflection from a semi-infinite step potential
become embedded in the contours of constant reduced action.$^{(\ref{bib:fpl13})}$

Let us now investigate the behavior of the reduced action for $A=1$ and $\beta = 0$
as a function of $B$. We still assume that $\hbar = 1$, $m=1$, and $k_x = k_y =
\pi/2$.  By symmetry, we can reduce the scope of our examination to a square in the
$x,y$-plane whose edges have one unit of length.  We present selected contours of
constant reduced action, $W_i$, for $\beta =0$ and various values of $-1 \le B \le
1$ on Fig.\ 5 (to reduce clutter on Fig.\ 5 and for consistency in this paragraph
with Fig.\ 5, we use positive and negative values of $B$ while restricting $\beta$
to zero). The straight contour in a diagonal direction of Fig.\ 5 for $B=0$ is
associated with the rectilinear propagation of the running wave function,
$A\exp[i(k_xx+k_yy)]$ without any interference. For $A=1$ and $B=1$, the straight,
piecewise continuous contours represent (cosine) standing-wave propagation and for
$B=-1$, (sine) standing-wave propagation. By Figs.\ 4 and 5, we examine the behavior
of the contours for reduced action as the value of $B$ changes from zero and
approaches the value of $1,-1$ for $A=1$ and $\beta = 0$.  The contours of constant
reduced action are initially straight for $B=0$. When the value of $B$ becomes
finite, interference exists. Then, as the finite value of $B$ increases toward the
value of $A=1$, the contours of constant reduced action become progressively more
serpentine. As $B \to 1$ for $\beta=0$, the radius of curvature goes to infinity for
all points on Fig.\ 5 except at $(x,y)=(0,-1)$ where the radius of curvature goes to
zero.  An analogous process exists for $B \to -1$ on Fig.\ 5. When $B \to \pm A$,
then the contours of constant reduced action become piecewise-continuous series of
square steps associated with standing-wave propagation.

\section{CONCLUSIONS}

We conclude that the trajectory representation including Faraggi and Matone's
quantum effective mass does describe interference without the need for Born's
quantum probability density.  In one dimension, a dichromatic particle contains all
the information regarding the two interfering monochromatic wave functions.  The
reduced action for the dichromatic particle is a complete solution to the QSHJE and
subsequently is the generator of motion for a solitary dichromatic particle which
itself is a solution to the time-independent Schr\"{o}dinger equation. A subsequent
trajectory can be developed for the interference patterns.

This represents a simplification.  First, as the dichromatic reduced action and the
reduced actions for the interfering wave functions are all complete solutions to the
QSHJE, each reduced action is described by the same number of nontrivial constants
of integration (as $\psi_d$ is entangled, the nontrivial constants of integration
for $W_1$ and $W_2$ may be redundant to some degree).  Only one set of non-trivial
constants of integration are needed to describe the motion of the interference
patterns.  Second, as the dichromatic reduced action is the generator of motion for
the interference pattern, one need only derive a single trajectory with its constant
coordinates to establish interference.

The trajectory representation describes the transition from running wave functions
through interfering wave functions to standing wave functions.  The trajectory
representation shows that interference induces quantum nonlocality where pairs of
interference patterns are created or annihilated. The segment of one member of the
pair associated with creation or annihilation manifests retrograde motion where
Faraggi and Matone's quantum effective mass is negative.

If a wave function for a free particle subject to nil potential is not
monochromatic, then time reversals and the attendant pair creation and annihilations
will eventually be induced.  Pair creations and annihilation of interference
patterns do not imply any high energy process as Faraggi and Matone's quantum
effective mass for the dichromatic wave function becomes nil at these points. The
trajectory representation provides a different interpretation into what Copenhagen
had heretofore considered as wave-packet spreading. The trajectory representation
also extends the concept of group velocity of wave packets beyond the integrity of
the wave packet.  It also generalizes the concept of group velocity for wave packets
to cases where spectral components may have significantly different amplitudes.

The trajectory for $\psi_d(x)$ renders insight into quantum nonlocality for
entangled spectral components, $\psi_1(x)$ and $\psi_2(x)$.  The time reversal of
the trajectories of $\psi_d(x)$ exhibit the nature of ``spooky action at a distance"
for entangled wave functions by inducing time reversals and retrograde motion.

\bigskip

\noindent {\bf Acknowledgements}

\bigskip

I heartily thank Marco Matone for his interesting discussions.  I also thank Robert
Carroll, Alon E. Faraggi, Bill Poirier and Robert E. Wyatt for their contributions
and encouragement.

\bigskip

\noindent {\bf References}

\begin{enumerate}\itemsep -.06in

\item \label{bib:prd34} E.\ R.\ Floyd, {\it Phys.\ Rev.}\ {\bf D 34}, 3246 (1986).

\item \label{bib:vigsym3} E.\ R.\ Floyd, ``The Philosophy of the Trajectory
Representation of Quantum Mechanics" in  {\it Gravitation and Cosomology: From the
Hubble Radius to the Planck Scale; Proceedings of a Symposium in Honour of the
80$^{th}$ Birthday of Jean-Pierre Vigier}, ed.\ by R.\ L.\ Amoroso, G.\ Hunter, M.\
Kafatos and J.-P.\ Vigier, (Kluwer Academic, Dordrecht, 2002) pp 401-408, extended
version promulgated as quant-ph/00009070.

\item \label{bib:fm} A.\ E.\ Faraggi and M.\ Matone, {\it Phys.\ Rev.\ Lett.}\ {\bf
78}, 163 (1997) hep-th/9606063; {\it Phys.\ Lett.}\ {\bf B 437}, 369 (1997),
hep-th/9711028; {\bf B 445}, 77 (1999), hep-th/9809125; 357 (1999), hep-th/9809126;
{\bf B 450}, 34 (1999), hep-th/9705108; {\bf A 249}, 180 (1998), hep-th/9801033.

\item \label{bib:fm2} A.\ E.\ Faraggi and M.\ Matone, {\it Int.\ J.\ Mod.\ Phys.}\
{\bf A 15}, 1869 (2000) hep-th/98090127.

\item \label{bib:bfm} G.\ Bertoldi, A.\ E.\ Faraggi and M.\ Matone, {\it Class.\
Quant.\ Grav.}\ {\bf 17} 3965 (2000), hep-th/9909201.

\item \label{bib:rc} R.\ Carroll, {\it Can.\ J.\ Phys.}\ {\bf 77}, 319 (1999),
quant-ph/9904081; {\it Quantum Theory, Deformation and Integrability} (Esevier,
2000, Amsterdam) pp. 50--56; {\it Uncertainty, Trajectories, and Duality},
quant-ph/0309023.

\item \label{bib:prd25} E.\ R.\ Floyd, {\it Phys.\ Rev.}\ {\bf D
25}, 1547 (1982).

\item \label{bib:prd26} E.\ R.\ Floyd, {\it Phys.\ Rev.}\ {\bf D
26}, 1339 (1982).

\item \label{bib:fpl9} E.\ R.\ Floyd, {\it Found.\ Phys.\ Lett.}\
{\bf 9}, 489 (1996), quant-ph/9707051.

\item \label{bib:prd29} E.\ R.\ Floyd {\it Phys.\ Rev}\ {\bf D 29}, 1842 (1984).

\item \label{bib:afb20} E.\ R.\ Floyd, {\it Ann.\ Fond.\ L.\ de
Broglie} {\bf 20}, 263, (1995).

\item \label{bib:misc} G.\ Reinisch, {\it Physica} {\bf A 206}, 229 (1994); {\it
Phys.\ Rev.}\ {\bf A 56}, 3409 (1997); M.\ Matone, {\it Found. Phys.\ Lett.\ } {\bf
15}, 311 (2002), hep-th/0005274; E.\ R.\ Floyd, {\it Int.\ J.\ Theor.\ Phys.}\ {\bf
27}, 273 (1988); {\it Phys.\ Lett.\ }{\bf A 214}, 259 (1996) {\bf A 206}, 307
(2002), quant-ph/0206114; {\it Int.\ J.\ Mod.\ Phys.\ }{\bf A 14}, 1111 (1999),
quant-ph/9708026; {\bf A 15}, 1363 (2000), quant-ph/9907092; M.\ V.\ John, {\it
Found.\ Phys.\ Lett.\ }{\bf 15}, 329 (2002), quant-ph/0109093.

\item \label{bib:bouda}  A.\ Bouda, {\it Found.\ Phys.\ Lett.}\ {\bf 14}, 17 (2001),
quant-ph/0004044; {\it Int.\ J.\ Mod.\ Phys.}\ {\bf A 18}, 3347 (2003),
quant-ph/0210193; A.\ Bouda and T. Djama, {\it Phys.\ Lett.}\ {\bf A 285}, 27
(2001), quant-ph/0103071; {\bf A 296}, 307 (2002), quant-ph/0206149; {\it Phys.\
Scripta} {\bf 66}, 97 (2002), quant-ph/0108022;

\item \label{bib:pe7} E.\ R.\ Floyd, {\it Phys.\ Essays} {\bf 7}, 135 (1994).

\item \label{bib:holland} P.\ R.\ Holland, {\it The Quantum Theory of Motion}
(Cambridge University Press, Cambridge, 1993) pp.\ 86--87, 141--146.

\item \label{bib:wyatt} R.\ E.\ Wyatt, {\it Quantum Dynamics with Trajectories:
Introduction to Quantum Hydrodynamics} (Springer, New York, 2005) Chap.\ 14.

\item \label{bib:ww} E.\ R.\ Floyd, {\it Found.\ Phys.}\ {\bf 37}, 1403 (2007),
quant-ph/0605121.

\item \label{bib:pe5} E.\ R.\ Floyd, {\it Phys.\ Essays} {\bf 5}, 135 (1992).

\item \label{bib:poirier} B.\ Poirier, {\it J.\ Chem.\ Phys.}\ {\bf 121}, 4501
(2004).

\item \label{bib:fpl13} E.\ R.\ Floyd, {\it Found.\ Phys.\ Lett.}\ {\bf 13}, 235
(2000), quant-ph/9708007.

\item \label{bib:barton} G.\ Barton, {\it Ann.\ Phys.}\ (N.Y.) {\bf 166}, 322 (1986).

\item \label{bib:hartman}  Hartman, {\it J.\ App.\ Phys.}\ {\bf 33}, 3247 (1962).

\item \label{bib:fletcher} Fletcher, {\it J.\ Phys.}\ {\bf C 18}, L55 (1985).

\item \label{bib:knobloch} A.\ S.\ Landsberg and  E.\ Knobloch, {\it Phys.\ Lett.}\
{\bf A 159}, 17 (1991); E.\ Knobloch, A.\ S.\ Landsberg, J.\ Moehlis, {\it Phys.\
Lett.}\ {\bf A 255}, 287 (1999); C.\ Martel, E.\ Knobloch, J.\ M.\ Vega, {\it
Physica} {\bf D}: {\it Nonlinear Phenomena} {\bf 137}, 94 (2000).

\item \label{bib:bw} M.\ Born and E. Wolf, {\it Principle of Optics} (Pergamon
Press, New York, 1959) pp 18--22.

\end{enumerate}

\bigskip

\noindent {\bf Figure Captions}

\bigskip

\noindent Fig.\ 1. Motion, $x(t)$, of the dichromatic wave function for $\tau=0$,
$A=1$, $B=0.5$, $k=\pi/2$ and $\beta = 0$ as a solid line and for $\beta = \pi$ as a
dashed line.  The turning points are a manifestation of nonlocality in quantum
mechanics.  At the turning points on the left where local minima of time exist,
pairs of maximum interference patterns are created.  One pattern propagates with
advancing motion; the other, in retrograde motion.  Likewise, at the turning points
on the right where local maxima of time exist, pairs of minimum interference
patterns merge where one pattern has an advancing trajectory while the other has a
retrograde trajectory.  As an interference pattern propagates along a trajectory
from a local minimum in time to a local maximum in time, the interference pattern
monotonically becomes more destructive.

\bigskip

\noindent Fig.\ 2. Motion of the dichromatic wave function, $x(t)$, for $\tau=0$,
$A=1$, $B=0.5$, $k=\pi/2$ and $\beta = 0,\pi/4,\pi/2, \cdots,7\pi/4$ for a set of
trajectories.  All trajectories are displayed as solid lines. With this spacing, the
left caustic formed by the set of trajectories near their turning points of local
minimum time is quite evident. The right caustic formed by the trajectories near
their turning points of local maximum time is not as evident with this degree of
trajectory spacing.  Only advancing and never retrograde segments of the
trajectories may become tangent to the caustics.

\bigskip

\noindent Fig.\ 3. Transformed effective quantum mass, ${\cal M}_{Q_d}$, and the
transformed velocity, $\stackrel{\bullet}{{\cal X}}_d$, as function of position,
$x$.   While ${\cal M}_{Q_d}$ is continuous, $\stackrel{\bullet}{{\cal X}}_d$ is
discontinuous at points where the velocity $\dot{x}$ jumps from $\pm \infty$ to $\mp
\infty$.  Retrograde motion is manifested by negative values of
$\stackrel{\bullet}{{\cal X}}_d$.

\bigskip

\noindent Fig.\ 4. Contours of constant reduced action and trajectory on the
$x,y$-plane for $y_0=0$, $A=1$, $B=0.5$, and $k_x=k_y=\pi/2$. The contours are
presented as solid lines with spacing between contours of $h/4$ units of action. A
trajectory launched from $(x,y)=(-2,-2)$ is presented as a dashed line.

\bigskip

\noindent Fig.\ 5.  Contours of constant reduced action on the $x,y$-plane for
$A=1$, $k_x=k_y=\pi/2$, $\beta = 0$, and $B=0,\pm 0.25,\pm 0.5,\pm 0.75,\pm 1$.  The
individual contours are designated by $B$.  The contour of constant reduced action
for $B=1$ is piecewise continuous consisting of two straight segments overlying the
edges of this Figure: one segment given by $y=0$; the other, $x=1$. And the contour
of constant reduced action for $B = -1$ is also piecewise continuous consisting of
two straight segments overlying the other edges of this Figure; one given by $x=0$;
the other, $y=-1$.

\end{document}